\documentclass[aps,pra,twocolumn,showpacs,amsmath,amsmath,superscriptaddress,amssymb,amsfonts]{revtex4-1}
\usepackage{graphicx}
\usepackage{dcolumn}
\usepackage{bm}
\usepackage{color}
\usepackage{multirow}
\usepackage{subfigure}
\usepackage{amsmath,amssymb}
\usepackage{braket}
\usepackage{latexsym}
\usepackage{amsfonts}
\usepackage[utf8]{inputenc}
\usepackage[colorlinks=true,citecolor=blue,linkcolor=blue]{hyperref}
\usepackage{ulem}

\begin{document}

\title{Quantum melting of two-component Rydberg crystals}

\author{Zhihao Lan}
\affiliation{School of Physics and Astronomy, University of Nottingham, Nottingham, NG7 2RD, UK}
\affiliation{
Centre for the Mathematics and Theoretical Physics of Quantum Non-equilibrium Systems,
University of Nottingham,
Nottingham NG7 2RD, UK
} 
\author{Weibin Li}
\affiliation{School of Physics and Astronomy, University of Nottingham, Nottingham, NG7 2RD, UK}
\affiliation{
Centre for the Mathematics and Theoretical Physics of Quantum Non-equilibrium Systems,
University of Nottingham,
Nottingham NG7 2RD, UK
} 
\author{Igor Lesanovsky}
\affiliation{School of Physics and Astronomy, University of Nottingham, Nottingham, NG7 2RD, UK}
\affiliation{
Centre for the Mathematics and Theoretical Physics of Quantum Non-equilibrium Systems,
University of Nottingham,
Nottingham NG7 2RD, UK
} 

\date{\today}
\begin{abstract}
We investigate the quantum melting of one dimensional crystals that are realized in an atomic lattice in which ground state atoms are laser excited to two Rydberg states. We focus on a regime where both, intra- and inter-state density-density interactions as well as coherent exchange interactions contribute. We determine stable crystalline phases in the classical limit and explore their melting under quantum fluctuations introduced by the excitation laser as well as two-body exchange. We find that within a specific parameter range quantum fluctuations introduced by the laser can give rise to a devil's staircase structure which one might associate with transitions in the classical limit. The melting through exchange interactions is shown to also proceed in a step-like fashion, in case of small crystals, due to the proliferation of Rydberg spinwaves.
\end{abstract}

\maketitle

{\it Introduction.}--- A long-standing topic in the study of condensed matter physics is the melting of low dimensional crystals that consist of interacting particles. In two dimensions (2D), it is widely accepted that thermally driven melting from a crystal to a liquid is a two-step procedure mediated by a hexatic phase according to the Kosterlitz, Thouless, Halperin, Nelson, and Young (KTHNY) scenario~\cite{KTHNY}. Interestingly, melting of quasi-1D crystals can proceed through either first or second order transitions, depending on the system parameters \cite{Levin_Dawson}. Both situations are different from 3D crystals which melt via a first order transition as predicted by Landau's mean field theory~\cite{3D_first_order}. Despite this broad understanding in the classical limit only little is known about the melting of crystals through quantum fluctuations.

In recent years there has been a growing effort to address the dimension-dependent crystallization and its melting by using  ultracold atomic and molecular gases. In 2D systems of cold polar molecules first-order superfluid-to-crystal transitions \cite{2Dmelt_molecule, 2Dmelt_molecule2} and the effect of quantum fluctuations on the formation of a hexatic phase \cite{Hexatic_zoller, Hexatic_bruun} have been theoretically investigated. In systems of Rydberg atoms crystalline phases \cite{Breyel12PRA, Pupillo10PRL, RydCry1,RydCry2,RydCry3,RydCry4,RydCry5,RydCry6,RydCry7,LanPRL15} and their melting \cite{2stage_melt, dislocation_melt, transport_melt} have attracted intensive attention and the experimental preparation of crystalline ground states (GSs) was reported \cite{RydCry_science} recently. The mechanism behind the quantum melting of a single-component Rydberg crystals in 1D is a two-stage process~\cite{2stage_melt} (similar to the KTHNY scenario), where a commensurate solid with true long-range order melts to a floating solid with quasi long-range order, and finally to a liquid phase.

 \begin{figure}[t]
\centering
\includegraphics[width=1.0\columnwidth]{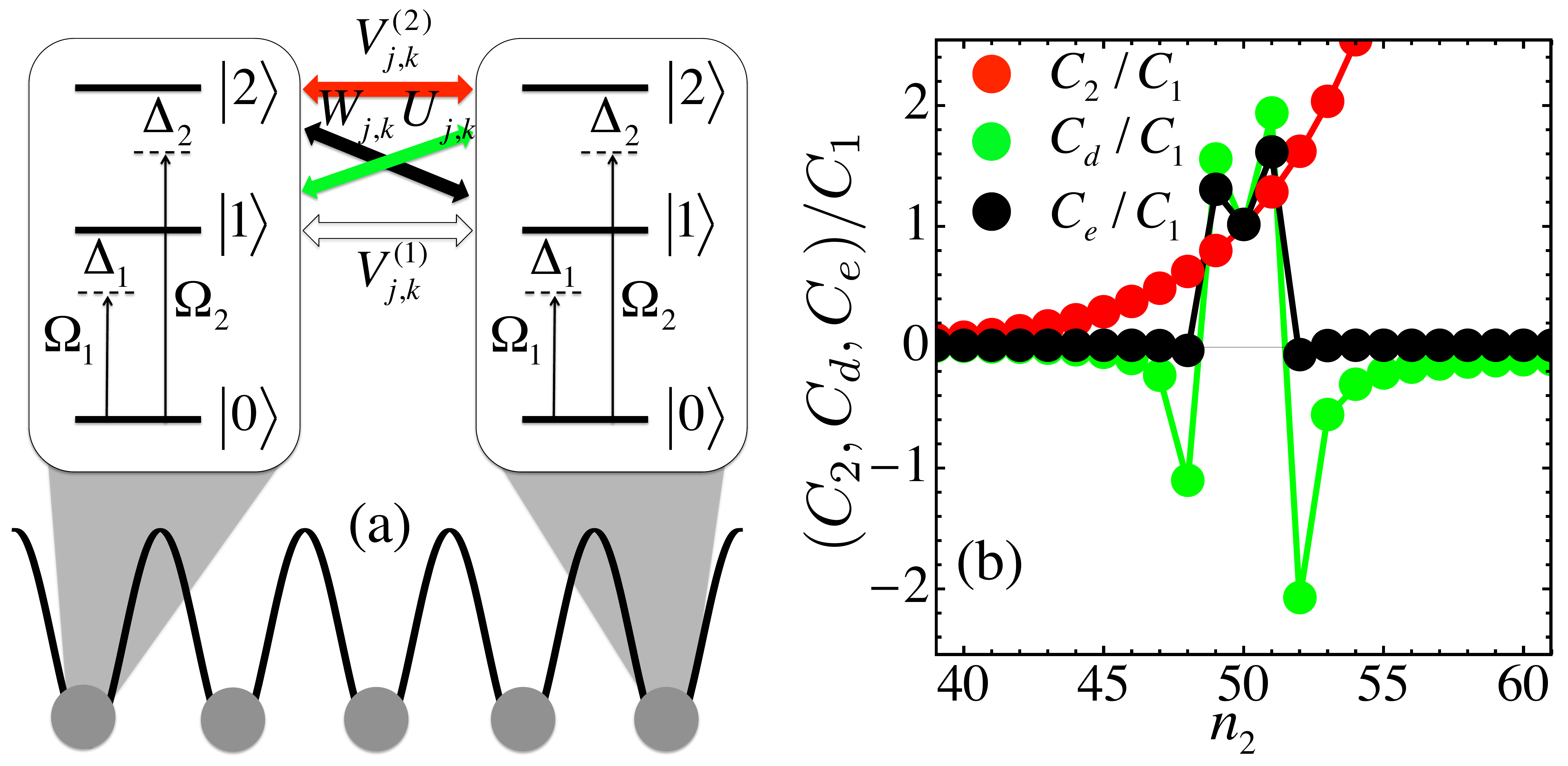}
\caption{(a) The system. Atoms are held in a 1D optical lattice. The atomic ground state $|0\rangle$ is laser excited to the Rydberg state $|1\rangle$ ($|2\rangle$) with Rabi frequency $\Omega_1$ ($\Omega_2$) and detuning $\Delta_1$ ($\Delta_2$). For two Rydberg atoms sitting at site $j$ and $k$, their intra-state, inter-state and exchange interaction are $V^{(\alpha)}_{j,k}$ ($\alpha=1,2$), $U_{j,k}$ and $W_{j,k}$, respectively. See text for details of the interaction potential. (b) Dispersion coefficients of the intra-state ($C_1, C_2$) and inter-state ($C_{\text{d}}$) vdW interaction as well as exchange interaction ($C_{\text{e}}$) for two Rydberg $S$-states of rubidium as a function of the principal quantum number $n_2$. We fix $n_1=50$, which yields $C_1=13.7$ GHz $\mu$m$^{6}$. Note the inter-state density-density vdW interaction $U_{j,k}$ and exchange interaction $W_{j,k}$ appear in the classical and quantum parts of the Hamiltonian respectively, see Eq.(\ref{Hamiltonian}).}
\label{fig1}
\end{figure}

The goal of this work is to shed light on melting mechanisms of 1D crystals in a physical setting in which two species of Rydberg atoms are excited. Such multi-component Rydberg gases currently receive much attention~\cite{excitation_transport1, Maxwell11PRL, Gunter13Science, Teixeira15PRL, Baur14PRL, Bettelli13PRA, Tiarks14PRL,Gorniaczyk14PRL,Fahey15PRA}. More importantly, the choice of this setting is that it permits the investigation of local and non-local quantum melting, driven by single and two-body processes, respectively. Atoms in Rydberg states experience strong van der Waals (vdW) type spin flip-flop (exchange) interactions, which can be comparable to their inter- and intra-state density-density vdW interaction~\cite{cooling1, cooling2, EIT_weibin}. Crystalline phases that are stabilized by the density-density interaction are melted by the laser coupling (local melting) and spin-exchange (non-local melting), respectively. In case of the local melting, the order parameter undergoes either a smooth or an abrupt (first order) transition. In the latter situation, the step-like structure resembles a devil's staircase that is typically observed in classical crystals~\cite{staircase} but not in the quantum regime. To shed light on the nonlocal melting process, we consider a parameter regime where only Rydberg states contribute to the many-body GS. Here the 1D Rydberg gas is described by the Heisenberg XXZ model. We demonstrate that a small Rydberg crystal is melted by the proliferation of delocalized Rydberg spinwaves, which also gives rise to discontinuous changes of the order parameter. Eventually, we identify specific configurations with which the quantum melting explored in this work can be realized experimentally with rubidium atoms.

{\it The System.}---  We consider atoms held in a 1D deep optical lattice (lattice spacing $d$ and number of lattice sites $L$) with one atom per site.  Each atom consists of three electronic states $|0\rangle,\, |1\rangle$, and $|2\rangle$. As shown in Fig.~\ref{fig1}(a), the atomic GS $|0\rangle$ is laser coupled to the Rydberg state $|1\rangle$ ($|2\rangle$) with Rabi frequency $\Omega_1$ ($\Omega_2$) and detuning $\Delta_1$ ($\Delta_2$). The detuning $\Delta_1$ ($\Delta_2$) effectively acts as a chemical potential for the  state $|1\rangle$ ($|2\rangle$). For two Rydberg atoms located on sites $j$ and $k$, we parameterize their intra-state and inter-state density-density interaction by  $V^{(\alpha)}_{j,k}=V_{\alpha}/(j-k)^6$ and $U_{j,k}=U/(j-k)^6$, and the exchange interaction by $W_{j,k}=W/(j-k)^6$, where
 $V_{\alpha}=C_{\alpha}/d^6$ ($\alpha=1,2$),  $U=C_{d}/d^6$ and $W=C_{e}/d^6$ denote the corresponding nearest-neighbour (NN) interactions. Here $C_{\alpha}$, $C_{\text{d}}$ and $C_{\text{e}}$ are the respective dispersion coefficients. This yields the following Hamiltonian for the system, which we write as the sum of a classical ($H_{\text{c}}$) and a quantum ($H_{\text{q}}$) term:
  \begin{gather}
 H=H_{\text{c}}+H_{\text{q}},  \label{Hamiltonian} \\
 H_{\text{c}}= \sum_{l>k,\alpha}\left[V_{\alpha}\frac{n^{(\alpha)}_k n^{(\alpha)}_l}{\left(l-k\right)^6}+\frac{U}{2}\sum_{\alpha'\neq \alpha}\frac{n^{(\alpha)}_k n^{(\alpha')}_l}{\left(l-k\right)^6}\right]-\sum_{k,\alpha}\Delta_{\alpha}n^{(\alpha)}_k,  \nonumber  \\
  H_{\text{q}}=\sum_{k}^{}\left[W\sum_{l>k}\frac{\sigma^{(+)}_k \sigma^{(-)}_l+\sigma^{(-)}_k \sigma^{(+)}_l}{\left(l-k\right)^6} + \Omega_{1}\sigma^{(1)}_k+ \Omega_{2}\sigma^{(2)}_k \right]. \nonumber
 \end{gather}
 \noindent
  The local operators on site $j$ are given by $n_j^{(\alpha)}=|\alpha\rangle_j \langle \alpha |$, $\sigma_j^{(\alpha)}=|\alpha\rangle_j \langle 0 |+|0\rangle_j \langle \alpha|$, $\sigma_j^{(+)}=|2\rangle_j \langle 1 |$,  $\sigma_j^{(-)}=|1\rangle_j \langle 2 |$, where $\alpha=1,2$ denotes the two Rydberg states. We denote $H_{\text{c}}$ as classical as it contains only diagonal operators $n_j^{(\alpha)}$ acting on the local single particle Hilbert spaces. The quantum part $H_{\text{q}}$ on the other hand contains the off-diagonal operators $\sigma_j^{(\alpha)}$, $\sigma_j^{(+)}$ and $\sigma_j^{(-)}$. There is a large flexibility in tuning laser parameters ($\Delta_1$, $\Delta_2$,  $\Omega_1$, $ \Omega_2$ ). The strength of the vdW interaction is fixed by the specific choice of Rydberg states (see discussion towards the end of the paper). For convenience, energies will be scaled with respect to the NN interaction $V_1$ in the following.

{\it Classical two-component Rydberg crystals.}--- In the following we will investigate the nature of the GS in the classical limit, $H_{\text{q}}=0$. Note that certain aspects of this have been addressed by some of us in previous works~\cite{LeviNJP15, LeviJSM16}, which were however limited to very specific parameter values, i.e., $V_2 \gg V_1$ and $U=W=0$. There it was shown that the presence of the strongly interacting species ($V_2$) can lead to frustration effects preventing the weakly interacting species ($V_1$) from assuming its lowest energy configuration.

To understand the coarse structure of the classical crystalline GS configurations, we will for the moment approximate the vdW interactions as NN interactions. Using the technique of {\it irreducible blocks} (see \cite{suppl_material} for an introduction to the technique and \cite{block} for the original reference), seven possible irreducible blocks $\{0,1,2,01,02,12, 012\}$) are identified, which provide the unit cell structure of GS crystals. Their energy densities can be found analytically and are summarized in Table \ref{table}. Note that the phase $\mathrm{VII}$ cannot be the GS of the system for any set of parameters, due to $E_{\mathrm{II}}+E_{\mathrm{IV}}+E_{\mathrm{VI}}=3E_{\mathrm{VII}}$, i.e. its energy is always larger than at least one of the other phases.

\begin{figure}[t]
\centering
\includegraphics[width=1.0\columnwidth]{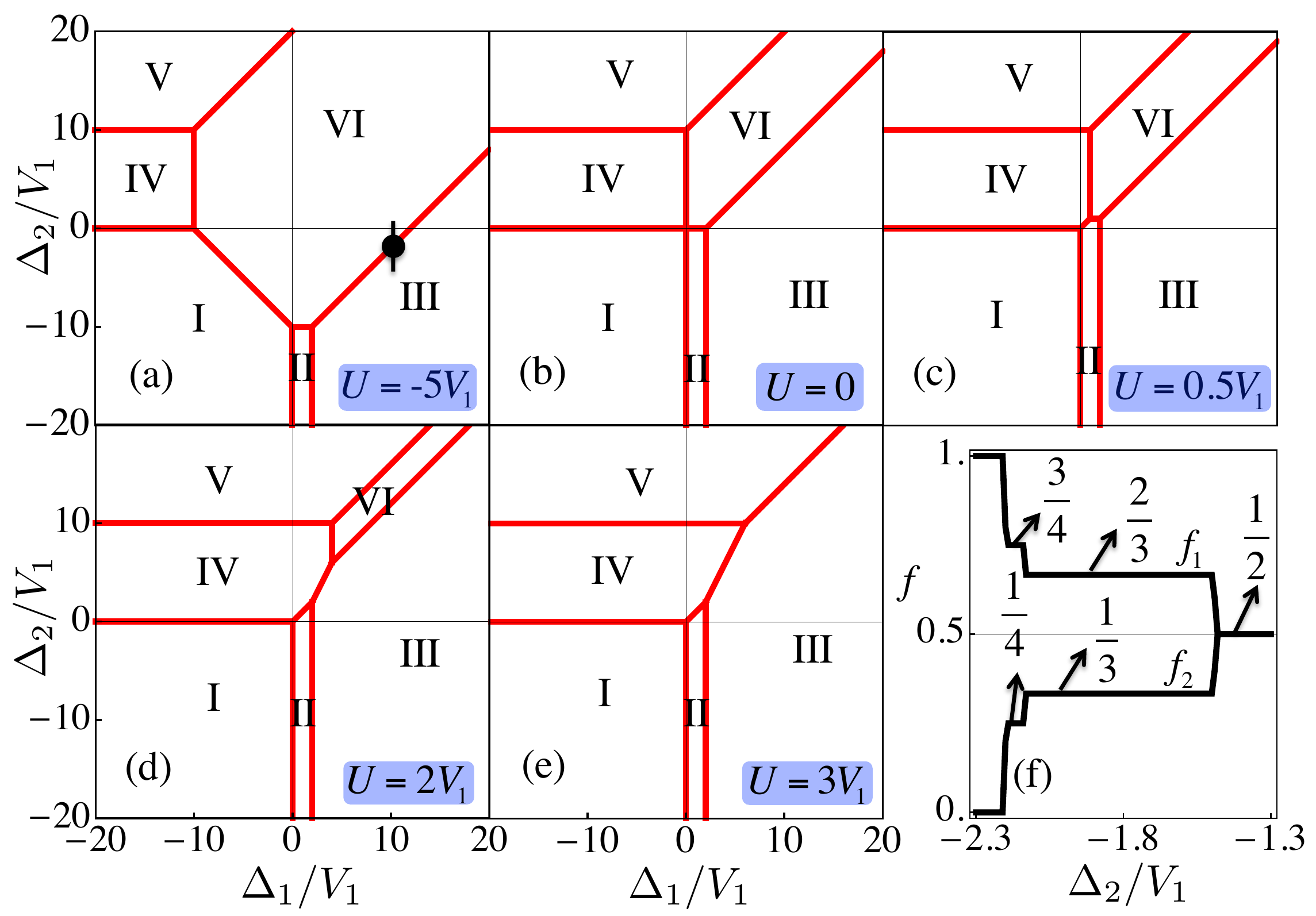}
\caption{Ground state phase diagrams of the classical Hamiltonian $H_{\text{c}}$ for $U=-5V_1$ (a),  $0$ (b), $0.5V_1$ (c), $2V_1$ (d), and $3V_1$ (e), obtained by explicitly checking which configuration of the crystalline phases - as listed in Table \ref{table} - has the lowest energy density for given $\Delta_{1,2}$. We consider here only the NN interaction and set $V_2=5V_1$. Panel (f) shows a magnification of the staircase structure in the vicinity of $(\Delta_1,\Delta_2)=(10,-2)V_1$ along $\Delta_2$ as marked in (a) that emerges when including the vdW tail. Here we display the populations, $f_{\alpha}=\sum_{j=1}^{L}n_j^{(\alpha)}/L$, of the atomic state $| \alpha\rangle$.}
\label{fig2}
\end{figure}
\begin{table}[ht]
\caption{ The seven possible crystalline phases of $H_{\text{c}}$ and their corresponding energy densities.}
\centering
\begin{tabular}{c c c}
\hline\hline
Label & Configuration  & Energy density \\ [0.5ex]
\hline
$\mathrm{I}$&000 $\cdots$ &$E_{\mathrm{I}}=0$ \\
$\mathrm{II}$&101010$\cdots$ &$E_{\mathrm{II}}=-\Delta_1/2$ \\
$\mathrm{III}$&111$\cdots$ &$E_{\mathrm{III}}=-\Delta_1+V_1 $ \\
$\mathrm{IV}$ &202020$\cdots$ &$ E_{\mathrm{IV}}=-\Delta_2/2 $ \\
$\mathrm{V}$ & 222$\cdots$ & $E_{\mathrm{V}}=-\Delta_2+V_2 $\\
$\mathrm{VI}$ & 121212$\cdots$ & $E_{\mathrm{VI}}=(-\Delta_1-\Delta_2+2U)/2 $\\
$\mathrm{VII} $ & 012012012$\cdots$ & $E_{\mathrm{VII}}=(-\Delta_1-\Delta_2+U)/3 $\\ [1ex]
\hline \hline
\end{tabular}
\label{table}
\end{table}

In Fig.\ref{fig2}(a-e), we present phase diagrams in the $\Delta_1-\Delta_2$ plane for different values of the inter-state interaction $U$. In each situation, the crystal configuration can be changed from one containing no Rydberg excitation, to a single-component or a two-component Rydberg crystal, by modifying the laser detuning $\Delta_1$ or $\Delta_2$. When comparing these panels, the relative areas occupied by different phases are modified by $U$. For examples, the region occupied by the composite crystalline phase $\mathrm{VI}$ first shrinks and finally disappears when $U$ increases from $-5V_1$ to $3V_1$ (the phase diagram no longer changes when $U\geq 3V_1$).

Let us now investigate the effect of the tail of the vdW interaction on the classical GS phase diagrams of Fig.\ref{fig2}(a-e). In 1D (single component) Ising models, it has been shown that such algebraically decaying potentials lead to the formation of a devil's staircase~\cite{staircase}. This is a fractal structure whose steps (or plateaus) are defined as the stability regions of configurations with specific rational filling fractions (density of excitations). Such structure is also formed in the two-component system in the vicinity of the phase boundaries displayed in Fig.\ref{fig2}(a-e). As an example, we calculate stable classical crystalline phases in the transition region between the phases $\mathrm{III}$ and $\mathrm{VI}$, around the point marked in Fig.~\ref{fig2}(a). The calculation is done by explicitly checking which rational filling fraction, of the form $f=p/q$ (with $p\leq q$ and maximal $q=13$) of an infinite system with period $q$, has the lowest energy per site ~\cite{LanPRL15}. Performing calculations with large $q>13$ makes the numerics more tedious and also adds little information to the coarse structure of the staircase as stable configurations with large $q$ normally correspond to high commensurate phases with very narrow steps. In Fig.~\ref{fig2}(f) we display the populations of the atomic states, $f_{\alpha}=\sum_{j=1}^{L}n_j^{(\alpha)}/L$, ($\alpha=0,1,2$) --- which in the following serve as {\it an order parameter} --- as a function of $\Delta_2$. We observe a number of steps --- reminiscent of a devil's staircase structure --- on each of which the components of order parameter assume rational values different from those corresponding to the phases of Table \ref{table}. Hence each plateau represents a new crystalline phase with narrower stability region. For example, the second largest plateau corresponds to $f_1=2/3$ (or $f_2=1/3$). Its length along the $\Delta_2$-axis is $0.63V_1$, which is only about $2\%$ of the phase $\mathrm{VI}$. An open question is whether our two-component system can indeed form a complete devil's staircase~\cite{staircase}.

{\it Laser induced local melting.}--- It was found that the laser induced melting of a single-component Rydberg crystal is a continuous and two-stage process~\cite{2stage_melt, dislocation_melt}. In contrast, we will illustrate here that such local melting of a two-component Rydberg crystal can proceed via a series of discontinuous transitions. To this end, we consider the case in which the exchange interaction between Rydberg states can be neglected. We begin by numerically diagonalizing a finite size system with $L=10$. The parameters of the laser driving the $|0\rangle$-$|1\rangle$-transition [see Fig. \ref{fig1}(a)] are fixed to $\Omega_1=0$ and $\Delta_1=10V_1$ such that the accessible classical phases are given by the configurations $\mathrm{III}$, $\mathrm{V}$ and $\mathrm{VI}$ [see Fig.\ref{fig2}(a-d)]. The crystal melting is then solely effectuated by the second laser whose Rabi frequency $\Omega_2$ we vary. With this particular set of parameters (i.e., $\Omega_1=0$ and varying $\Omega_2$), atoms in state $|1\rangle$ remain essentially ``classical" while the states $|2\rangle$ and $|0\rangle$ form a superposition that ultimately leads to the quantum melting of classical crystalline states. We will discuss the effect of finite $\Omega_1$ on the melting process further below.

 \begin{figure}[t]
\centering
\includegraphics[width=1.0\columnwidth]{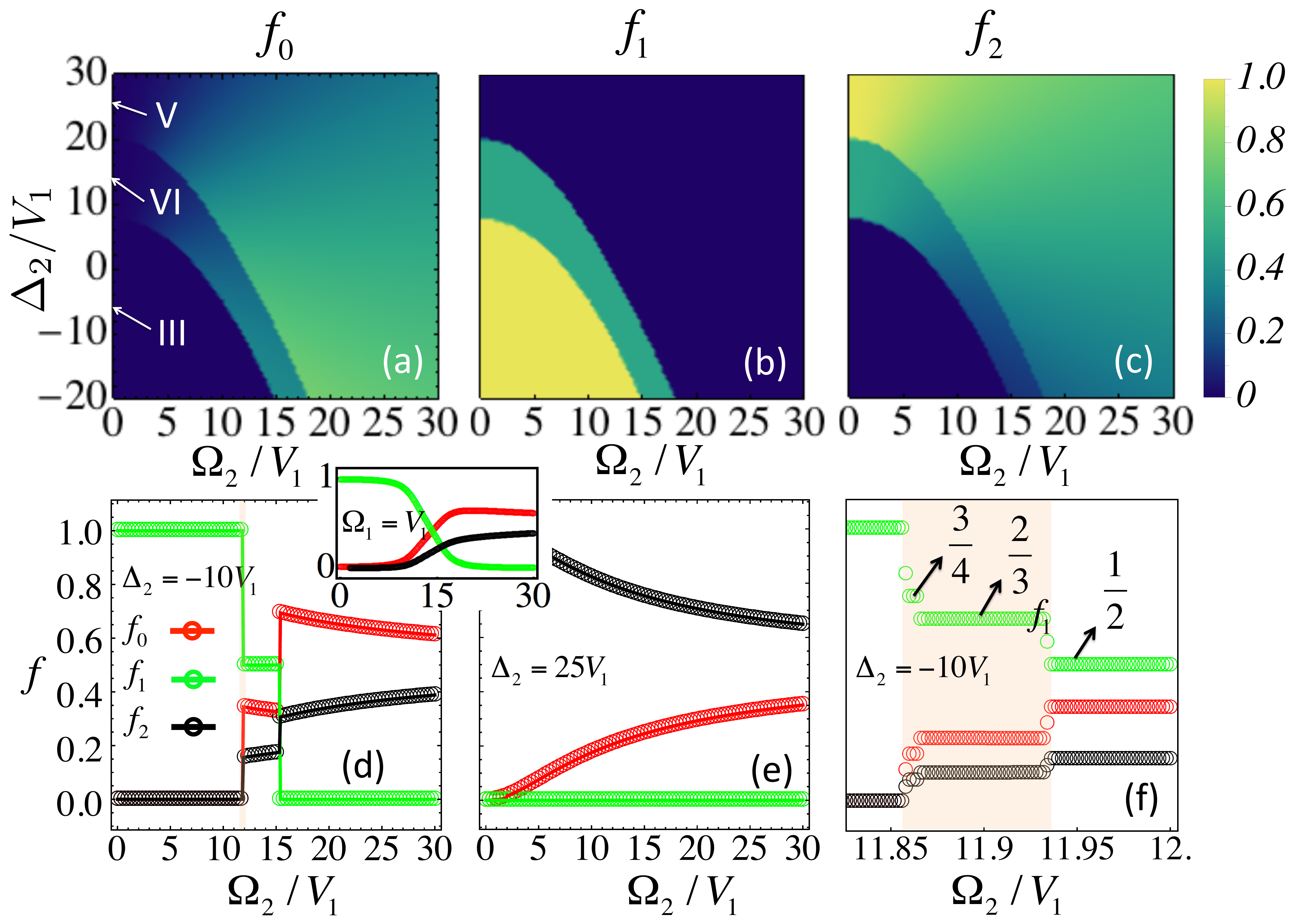}
\caption{Laser induced local melting with $\Omega_1=0$  while varying $\Omega_2$. (a-c) Populations $f_0$, $f_1$ and $f_2$. The sharp change of populations clearly demonstrates the first-order nature of the laser induced melting (in the classical limit the phases $\mathrm{III}$, $\mathrm{V}$ and $\mathrm{VI}$ are marked by arrows). The considered Rydberg states are $n_1=50$ and $n_2=57$ (such that $W\simeq 0$, see Fig. \ref{fig1}) and $\Delta_1=10V_1$. The data is obtained by the exact diagonalization of a finite lattice with $L=10$ under periodic boundary conditions, where the tail of the vdW interaction is included in the numerics. In panel (d), the first order transition is clearly visible. In panel (e), a smooth melting transition is shown. The mean-field results (solid lines) with only NN interactions agree well with the numerical diagonalization results (circles). (f) Magnification of the devil's staircase in the vicinity of $\Omega_2=11.9V_1$ as shown in panel (d). To resolve next largest plateaus at $f_1=2/3$ and $3/4$, a larger lattice with $L=12$ was chosen for the calculation. Note that the staircase structure vanishes for finite values of $\Omega_1$ as can be seen in the inset of panel (d). Here $\Omega_1=V_1$ with all other parameters being the same as in (d).}
\label{fig3}
\end{figure}

The components $f_{\alpha}$ of the order parameter are shown in Fig.~\ref{fig3}(a-c). Additional cuts along $\Delta_2=-10V_1$ are provided in Fig.~\ref{fig3}(d). The data indicates a number of sharp jumps reminiscent of {\it first order} transitions that start from the classical limit ($\Omega_2=0$) and extend into the quantum regime. For example, in Fig.~\ref{fig3}(d), when $\Omega_2$ is smaller than a critical value $\Omega_{\rm L}\approx 11.9V_1$, the GS is formed by atoms in state $|1\rangle$ --- the phase $\mathrm{III}$ --- and the laser $\Omega_2$ in fact has no effect. However, once $\Omega_2>\Omega_{\text{L}}$, all three atomic states are populated suddenly, such that $f_1=0.5$ remains constant while the other two vary smoothly with respect to $\Omega_2$. By further increasing $\Omega_2$ one reaches a second critical value $\Omega_{\text{H}}\approx 15V_1$, from which onwards the population of $|1\rangle$ is completely suppressed and $f_0$ and $f_2$ change smoothly. Contrary to the above situation, melting of the $\mathrm{V}$ phase [Fig.~\ref{fig3}(e)] proceeds smoothly since this corresponds to the melting of a single-component Rydberg crystal~\cite{2stage_melt} that only involves the states $|2\rangle$ and $|0\rangle$.

The observed phase diagram is largely captured by a mean field (MF) theory where we write the site-decoupled GS wave function as $|\Psi\rangle=\prod_i \otimes \left(a_i |0\rangle_i+b_i |1\rangle_i+c_i |2\rangle_i\right)$~\cite{mft}. To illustrate the main mechanism we will again consider for the moment only NN interactions and as the unit cell occupies at most two sites with only NN interactions (see Table \ref{table}), the period of the wave function is two sites. The order parameter obtained from the MF calculation is in very good agreement with the diagonalization results [see Fig.~\ref{fig3}(d-e)]. MF further corroborates the first order nature of the observed transitions: when $0<\Omega_2<\Omega_{\text{L}}$, the wave function of a unit cell is given by a simple Fock state $|\psi_{A}\rangle=|11\rangle$. However, the wave function becomes  $|\psi_{B}\rangle=\alpha|10\rangle + \beta|12\rangle$ ($\alpha$ and $\beta$ are normalisation constants) when $\Omega_{\text{L}}<\Omega_2<\Omega_{\text{H}}$. The order parameter jumps at $\Omega_{\text{L}}$ as the two wave functions cannot be smoothly connected by merely varying $\alpha$ and $\beta$. This also highlights the nature of the first order transitions driven by $\Omega_2$: the $\Omega_2$-term of the Hamiltonian is minimized by a superposition of states $|0\rangle$ and $|2\rangle$. Increasing $\Omega_2$ (across $\Omega_{\text{L}}$) makes phase $\textrm{III}$ energetically unfavourable and leads to a partially crystalline phase. Here one of every two sites is occupied by atoms in state $|1\rangle$ and the other one is in a superposition of states $|0\rangle$ and $|2\rangle$. This is clearly different from the first order transitions observed in the classical limit where no superposition happens. Consequently, this partially crystalline phase features both crystalline antiferromagnetic correlations for state $|1\rangle$ and exponentially decaying density-density correlations for state $|2\rangle$.

Though driving by a quantum term $\Omega_2$, the tail of the vdW interaction leads to the emergence of a devil's staircase in the vicinity of the transition points for state $|1\rangle$, which behaves classically as $\Omega_1=0$. The corresponding numerical data around $\Omega_2=\Omega_{\text{L}}$ is shown in Fig.~\ref{fig3}(f). Here multiple plateaus emerge between the two main plateaus corresponding to $f_1=1$ and $f_1=0.5$. Transitions between plateaus proceed similarly to the discussion above: on each plateau atoms in the state $|1\rangle$ form a crystalline structure, whose staircase has the same pattern as its classical counterpart [see Fig.~\ref{fig2}(f)]. However the sites that were originally occupied by an atom in state $|2\rangle$ now enter a superposition state and ``melt".  We would like to point out that the staircase of $f_1$ displayed in Fig. \ref{fig3}(f), exhibits the same plateaus as its classical counterpart given in Fig. \ref{fig2}(f). The steps in the population are thus physical and a consequence of the ``classical species" (in state $\left|1\right>$) adapting its density in order to achieve the overall lowest energy state of the system. Quantum fluctuations introduced by a finite coupling $\Omega_1$ smear out the staircase. This is shown in the inset of Fig.\ref{fig3}(d).

{\it Exchange interaction induced nonlocal melting.}---To discuss the non-local melting we consider a regime where only the two Rydberg states play roles in the physics. This is achieved when  $\Omega_{\alpha}=0$ and $\Delta_{\alpha}$ ($\alpha=1,2$) is sufficiently large, such that classically the GS can only be one of the phases $\mathrm{III}$, $\mathrm{V}$ and $\mathrm{VI}$. With this choice of parameters, the state $|0\rangle$ is never populated even when the many-body GS is away from the classical limit.

\begin{figure}[t]
\centering
\includegraphics[width=1.0\columnwidth]{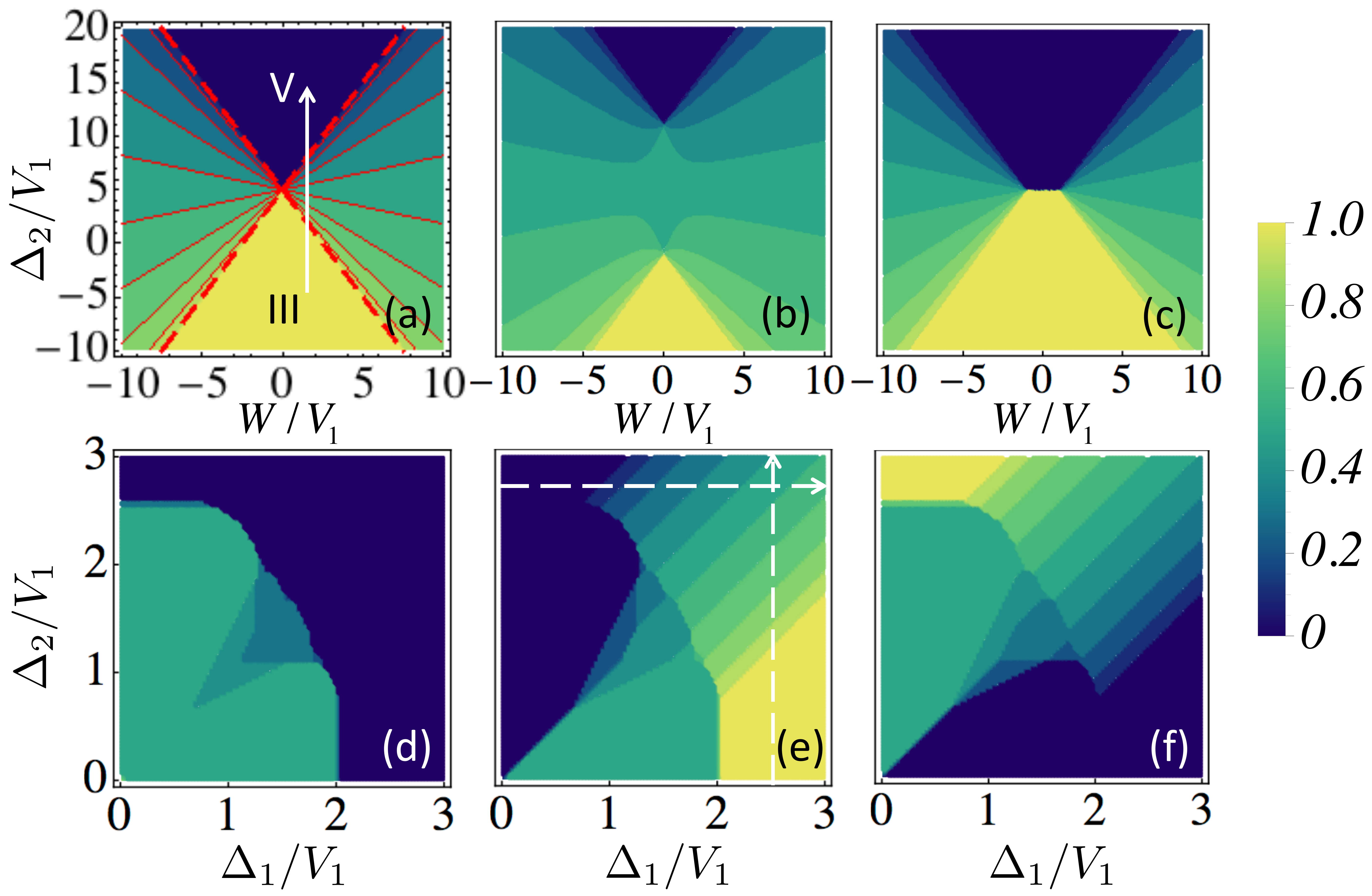}
\caption{ Population $f_1$ for $U=V_1$ (a), $-2V_1$ (b), and $2V_1$ (c). The red (solid and dashed) lines in (a) are the analytic results with only NN interactions from Eq. (\ref{xxE}). Other parameters are $V_2=V_1$, $\Omega_1=\Omega_2=0$, and $\Delta_1=5V_1$. (d)-(f) Populations $f_0$, $f_1$, and $f_2$ for the experimental relevant case:  two Rydberg $S$-states of rubidium with principal quantum numbers $n_1=50$ and $n_2=51$ [see data in Fig.~\ref{fig1}(b)]. The melting of the phase $\textrm{III}$ ($\textrm{V}$) can be probed by changing $\Delta_2$ ($\Delta_1$) along the vertical (horizontal) arrow in (e). The tail of the vdW interaction is included in the numerical diagonalizations.}
\label{fig4}
\end{figure}

First we focus on a simplified situation in which the three relevant  interactions are of equal strength, i.e. $V_2=U=V_1$. By numerically diagonalizing the Hamiltonian (\ref{Hamiltonian}), we obtain the GS phase diagram of a small crystal of $L=10$. According to the population $f_1$ plotted in Fig.~\ref{fig4}(a), the system is in the crystalline phase $\mathrm{III}$ ($\mathrm{V}$)  when $\Delta_2$ is negative (positive) and $|\Delta_2|\gg W$. From the crystalline phase, $f_1$ jumps abruptly when we scan either $W$ or $\Delta_2$. For example, when increasing $\Delta_2$ along the vertical arrow shown in Fig.~\ref{fig4}(a), the phase $\mathrm{III}$ melts at the first jump and the new many-body GS contains one more excitation in state $|2\rangle$. This process repeats at every jump until  $f_2=1$ ($f_1=0$), i.e., the phase $\mathrm{V}$.

To understand this melting pattern, we project Hamiltonian (\ref{Hamiltonian}) to the subspace of the two Rydberg states and consider only NN interactions for simplicity. This reduces the system to a spin $1/2$ Heisenberg XX model with a field along the $\sigma_z$-direction,
\begin{equation}
H_{\text{XX}}=\sum_{i} \frac{W}{2} (\sigma^{x}_i \sigma^{x}_{i+1}+\sigma^{y}_i \sigma^{y}_{i+1})+h\sigma^{z}_i+C,
\label{xxH}
\end{equation}
where $h=(-V_1+V_2+\Delta_1-\Delta_2)/2$, $C=(V_1+V_2+2U-2\Delta_1-2\Delta_2)L/4$ and $\sigma^{\xi}_i\, (\xi=x,y,z)$ are the Pauli matrices for the two Rydberg states on site $i$.

This Hamiltonian can be analytically solved which permits to show that the melting of the phase $\mathrm{III}$ ($\mathrm{V}$) is due to a proliferation of Rydberg spinwave states. To be concrete, we will focus in the following on the melting of the phase $\textrm{III}$, whose wave function is given by $|\Psi_{\textrm{G}}\rangle=\Pi_k^L\otimes|1\rangle_k$. The eigenstates $|\Psi_{N}\rangle$ of Eq.~(\ref{xxH}) that contain a fixed number $N$ of spin excitations in state $|2\rangle$ can be explicitly calculated. For example for $N=1$, $|\Psi_{1}\rangle=1/\sqrt{L}\sum_j\sigma_j^{(+)}|\Psi_{\textrm{G}}\rangle$, is a spinwave where the single excitation in state $|2\rangle$ is shared by all the atoms in the lattice. From the eigenenergies $E_{N}^{\text{min}} = V_1-\Delta_1(L-N)/L-\Delta_2N/L-2W\sin(N\pi /L)/[L\sin(\pi /L)]$~\cite{xx_solution}, we obtain the transition from $N$ to $N+1$ excitations by varying the detuning $\Delta_2$,
\begin{gather}
\Delta_2=\Delta_1-\frac{2W}{\sin(\pi/L)}\left[\sin\frac{(N+1)\pi}{L}-\sin\frac{N\pi}{L}\right].
\label{xxE}
\end{gather}
These steps (see red solid and dashed lines in Fig. \ref{fig4}(a)) agree well with the position steps that were found in the numerics. The analytical results indicate that the crystal phase $|\Psi_{\textrm{G}}\rangle$ (i.e. phase $\textrm{III}$) switches to the delocalized spinwave state $|\Psi_{1}\rangle$ when we increase $\Delta_2$ (fixing $W$). The transition points, determined by $\Delta_2=\Delta_1-2W$, are highlighted by the two dashed lines in Fig.~\ref{fig4}(a). Note, that the step-like structure appears only for small sizes which are in fact relevant for current experiments~\cite{RydCry_science}. For macroscopic sizes the energy gaps between spinwave states vanish and the excitation density will vary continuously as a function of $W$.

Away from the special point $V_2=U=V_1$, the system is described by a Heisenberg XXZ model, $H_{\text{XXZ}}=H_{\text{XX}}+V_z \sum_{i} \sigma^{z}_i \sigma^{z}_{i+1}$ with $V_z=(V_1+V_2-2U)/4$, whose engineering in controllable quantum systems has attracted increased attention recently~\cite{xxz_cavity_epl, xxz_ions, xxz_molecule, xxz_p, xxz_Ryd12, xxz_Ryd1}. Here the presence of the $\sigma^{z}_i \sigma^{z}_{i+1}$-interaction terms changes the phase diagram structure. Two examples with $U=\pm 2V_1$ and $V_2=V_1$ are shown in Fig.~\ref{fig4}(b-c). Although the phase boundary changes, the melting of the crystalline phase $\mathrm{III}$ ($\mathrm{V}$) also proceeds through the proliferation of spinwave excitations, which has been verified by analyzing both the Hamiltonian (\ref{Hamiltonian}) and the effective Hamiltonian $H_\mathrm{XXZ}$.

{\it Experimental implementation of the quantum melting.}---Local melting is induced by controlling the excitation strength of Rydberg states. This has been realized in optical lattices or microtraps by several experimental groups \cite{Weber15NP, excitation_transport1, Labuhn14PRA, Maller15PRA, Viteau11PRL, Anderson11PRL,  RydCry_nature, RydCry_science, Beguin13PRL, Li13Nature, Urban09NP, Gaetan09NP}. In the following, we will focus on how to realize the nonlocal melting, which solely depends on the presence of two-body exchange interactions. One possible way to establish strong exchange interactions is to choose two Rydberg $S$-states whose principal quantum numbers $n_{\alpha}$ differ by $1$~\cite{EIT_weibin}. For example, dispersion coefficients for rubidium and $n_1=50$ and $n_2=51$ are $C_1=13.7$ GHz $\mu$m$^{6}$, $C_2=17.4$ GHz $\mu$m$^{6}$, $C_d=26.4$ GHz $\mu$m$^{6}$, and $C_e=21.9$ GHz $\mu$m$^{6}$. Alternatively, one could utilize the so-called F\"orster resonance to generate strong exchange interaction. In this case one can even tune the two-body interaction from a van der Waals to dipolar type with external electric fields~\cite{gorniaczyk}.

In the following, we will illustrate how to observe the nonlocal melting by using an example with the Rydberg $50S$ and $51S$ states. For lattice spacing $d=3\,\mu$m~\cite{Viteau11PRL}, we obtain a NN interaction of $V_1\approx 18.8$ MHz. Since the two-body interactions are fixed the non-local melting can be studied by changing the laser detunings $\Delta_1$ and $\Delta_2$. In Fig. \ref{fig4} (d-f), we present populations $f_\alpha$ of the state $|0\rangle$, $|1\rangle$ and $|2\rangle$ calculated with these parameters. Note, that compared to the ideal situation shown in Fig.~\ref{fig4} (a-c), the state $|0\rangle$ is in fact populated in certain parameter region [see lower-left corner in Fig.~\ref{fig4}(d)]. To probe the melting through spinwave proliferation of the Rydberg state, we have to avoid this parameter region. For example, one finds that $f_1=1$ when $\Delta_1=2.5V_1$ and $\Delta_2=0$. From here, we can then observe the melting of the phase $\mathrm{III}$ by increasing $\Delta_2$ as indicated by the vertical arrow in Fig.~\ref{fig4}(e) [See also Fig.~\ref{fig4}(a)].

{\it Outlook.}--- The goal of our study was to shed light on the nature of multi-component Rydberg crystals and in particular their melting under different kinds of quantum fluctuations. We found that, surprisingly, the quantum melting can proceed via first order phase transitions through a sequence of steps on a devil's staircase. The second melting mechanism, which proceeds through the proliferation of spinwaves, could potentially be employed for the deterministic creation of single- and multi-photon states~\cite{photon1,photon2,dudin12}.

\begin{acknowledgments}
\textit{Acknowledgements.}--- We thank R. M. W. van Bijnen, R. Nath and T. Pohl for discussions, and M. Marcuzzi for comments on the manuscript.
The research leading to these results has received funding from the
European Research Council under the European Union's Seventh Framework
Programme (FP/2007-2013) / ERC Grant Agreement No. 335266 (ESCQUMA), the
EU-FET Grant No. 512862 (HAIRS), the H2020-FETPROACT-2014 Grant No.
640378 (RYSQ), and EPSRC Grant No. EP/M014266/1. W.L. is supported through
the Nottingham Research Fellowship by the University of Nottingham.

\end{acknowledgments}

\newpage
\begin{widetext}
\section{Supplementary Material}
In this supplemental material, we give a brief introduction to the method of {\it irreducible blocks} used in the main text for finding the ground state phase diagrams of classical two-component Rydberg lattice gases. We will mainly focus on the concepts and how the method works. The details of the method can be  found in~\cite{block}. The key idea of the method is that for a long chain with finite range interactions, the total energy of the chain can always be written as a sum of energies of a set of basic blocks.  Now if one of the blocks has the lowest energy per site compared to other blocks, the whole chain tends to ``condense" to that block such that  the total energy of the chain is minimised and the ground state configuration of the chain would be periodic repetition of the block that has the lowest energy per site. In the following, we give more details. 

\begin{figure}[!h]
\centering
\includegraphics[width=0.7\columnwidth]{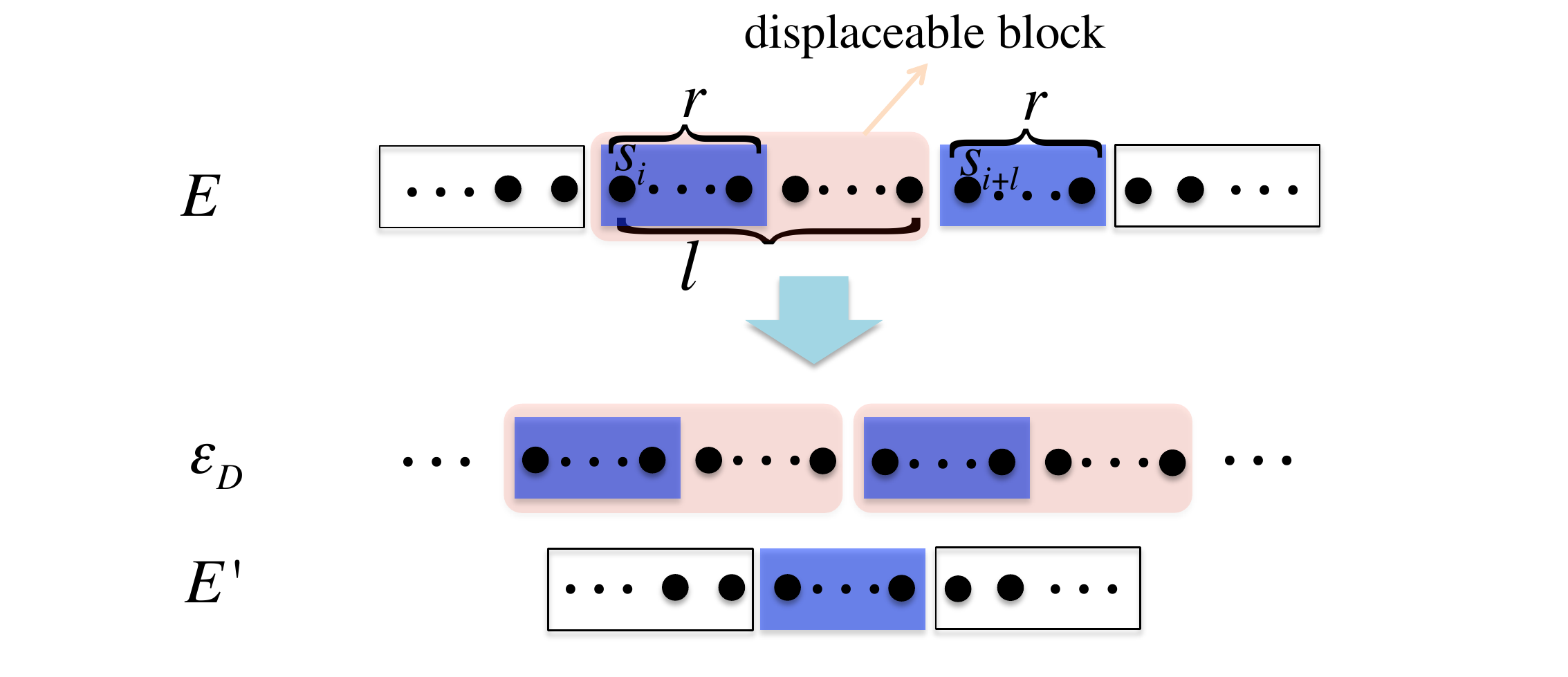}
\caption{(color online) The energy $E$ of a chain  with interactions of finite range $r$ can be written as the sum of the energy $E'$ of another chain obtained by removing a displaceable block $D$ and the energy per block $\epsilon_D$ of an infinite chain consisting of periodic repetition of that displaceable block. See text for details on the definition of displaceable block. }
\label{s1} 
\end{figure}

Suppose we have a classical one-dimensional lattice system with $L$ lattice sites, where each site $s_i$ can take $\sigma$ $ (0,1,2,\cdots,\sigma-1)$ possible configurations and the interaction of the system has a finite range of $r$. We first define a {\it displaceable block}  with $l$ sites from $i$ to $i+l-1$ (see Fig. (\ref{s1})) as following: if the configuration of the $r$ sites starting from $i$ and the $r$ sites from $i+l$ are isomorphic, i.e., if 

\begin{gather}
s_{i+j}=s_{i+l+j}, \hspace{2cm} 0\leq j\leq r-1
\end{gather}
we call the block of $l$ sites from $i$ to $i+l-1$ a {\it displaceable block}. If a chain involves a displaceable block, we call the chain reducible. Otherwise, we call it irreducible. Now if we remove the displaceable block from the chain, we find the energy $E$ of the original chain can be written as 

\begin{gather}
E=\epsilon_D+E'
\label{eq1}
\end{gather}
where $\epsilon_D$ is the energy per block of an infinite chain consisting of periodic repetition of the displaceable block and $E'$ denotes the energy of the chain obtained after removing the displaceable block from the chain.  
The very neat expression of (\ref{eq1}) comes from the fact that for interactions with a finite range $r$, the interaction terms around the boundary site of $s_i$ can be completely transferred to that around  site $s_{i+l}$ since the $r$ sites starting from $i$ and the $r$ sites starting from $i+l$ are exactly the same by the definition of displaceable block.  Now we define the reducibility or irreducibility of a block by the corresponding reducibility or irreducibility of the chain obtained from periodic repetition of the block, i.e., an {\it irreducible block} $B$ is a block such that there is no displaceable block in the infinite  chain obtained by periodic repetition of $B$.   So we can repeat the above process to remove further more displaceable blocks and at the end, we get 
\begin{gather}
E=\sum_B n_B \epsilon_B+E''
\end{gather}
where $E''$ is the energy of an irreducible chain and the sum is over all the irreducible blocks. Now it is clear, if $L$ is very large and if the minimum value $\epsilon_B/\nu_B$ (i.e., the energy per site of an irreducible block $B$ with number of lattice sites $\nu_B$), occurs for only one block, then the ground state configuration of the chain corresponds to the periodic repetition of that block, i.e., the ground state tends to ``condense" to the block that has the smallest energy per site. If the minimum value $\epsilon_B/\nu_B$ occurs for two or more blocks, then the ground state configuration can be a mixture of these blocks. 

To show how the above method works, we consider some examples. For $\sigma=2$, i.e., with two possible configurations (0 and 1) on each lattice site,  the irreducible blocks with different range interactions of $r=1,2,3$ are listed below (note we only retain the configurations that are invariant under rotation and/or refection, e.g., we consider 01 and 10 equivalent). 

\begin{itemize}
\item r=1: $\{ 0,1,01\}$
\item r=2: $\{ 0,1,01,010,011,0110\}$
\item  r=3: $\{ 0,1,01,010,011,0110,0100,01100,0111,01110,011100,010110,0101100,0101110,01011100\}$
\end{itemize}
For the Ising model with the nearest-neighbor interaction, $H_1=J\sum_{i}s_is_{i+1}-h\sum_i s_i$ with $s_i=(\uparrow, \downarrow)$ or $(+1,-1$) and $J>0$, we find the energy per site, $E_{\uparrow}=J-h$, $E_{\downarrow}=J+h$ and $E_{\uparrow\downarrow}=-J$ (we can only consider $h>0$, since the Hamiltonian is invariant under $s_i\rightarrow -s_i$ and $h\rightarrow -h$). So we find when $h>2J$, the ground state is ferromagnetic {$\uparrow\uparrow\uparrow\uparrow\cdots$} and when $h<2J$ the ground state is antiferromagnetic $\uparrow\downarrow\uparrow\downarrow\cdots$.  When the interaction range is $r=2$ and $r=3$, the corresponding Hamiltonian is $H_2=J_1\sum_{i}s_is_{i+1}+J_2\sum_{i}s_is_{i+2}-h\sum_i s_i$ and $H_3=J_1\sum_{i}s_is_{i+1}+J_2\sum_{i}s_is_{i+2}+J_3\sum_{i}s_is_{i+3}-h\sum_i s_i$. We can calculate the energy per site of all the irreducible blocks similarly. Thus we obtain the phase diagrams in the corresponding parameter space. 

For $\sigma=3$, where each lattice site can be in any of the three states $(0,1,2)$ as considered in the main text, we have 7 possible irreducible blocks when $r=1$,  
\begin{itemize}
\item r=1: $\{ 0,1,2,01,02,12,012\}$
\end{itemize}

The energies per site of the seven blocks of the two-component classical Rydberg lattice gas with only nearest-neighbour interaction (i.e., $r=1$) have been given in Table 1 of the main text (note the rule for $r=1$ is that the irreducible block can not have two same onsite configurations, otherwise the size of the block can be reduced by the definition of irreducible blocks). By investigating which block has the lowest energy density we have obtained the ground state phase diagrams as presented in Figure 2 of the main text. 

Though in principle, the method of irreducible blocks can be applied to any one-dimensional classical lattice models with any finite range interactions, in practice, the number of irreducible blocks increases very rapidly with the range  $r$ of the interactions, for example, for $r=2$ and three states each site, there are 87 irreducible blocks in total \cite{block}.  When $r\rightarrow \infty$, there are infinite possible ground states. This leads to  the complete devil's staircase of long-range interacting Ising models \cite{staircase}. 

\end{widetext}

\end{document}